\documentclass[12pt]{article}
\usepackage{epsfig}
\usepackage{graphics}
\usepackage{graphicx}
\usepackage[centerlast,footnotesize]{caption2}
\begin{document}

\begin{center}
{\bf \Large  Production of $\Omega_{scb}$ Baryons in Electron-Positron 
Collisions}

\vspace{1.12cm}
S. P. Baranov$^{\it a}$\footnote{e-mail: baranov@sci.lebedev.ru} and 
V. L. Slad$^{\it b}$\footnote{e-mail: vslad@theory.sinp.msu.ru}

\vspace{0.75cm}

$^{\it a}${\small Lebedev Institute of Physics, Russian Academy of Sciences, 
Leninskii pr. 53, Moscow 119991, Russia}

$^{\it b}${\small Skobeltsyn Institute of Nuclear Physics, Moscow State 
University, Moscow 119992, Russia}
   
\vspace{1.35cm}                          

{\bf Abstract}
\end{center}\vspace*{-0.35cm}
{\small The total and differential cross sections for the production of 
$\Omega_{scb}$ baryons in electron-positron collisions are calculated at the 
$Z$-boson pole.}

\vspace{1.02cm}

{\bf \center \Large 1. Introduction}

Baryons involving two or three heavy quarks ($c$, $b$) have not yet been 
observed experimentally. Theoretical investigations into the mass spectra of 
hadrons that contain two or more heavy quarks, the cross sections for their 
production in various processes, and their lifetimes and decay modes form a 
rather new line of research in particle physics. For an overview of these 
investigations, the interested reader is referred to [1]. Calculations 
presented in the literature that are aimed at determining the production cross 
sections for baryons featuring two heavy quarks either rely on the 
fragmentation approach [2], or treat the production of unbound four-quark 
states in a bounded phase space [3], or consider (as is done in the majority of
the most recent studies on this subject) the production of the relevant heavy 
diquark [4-9]. In particular, the production of doubly heavy baryons in 
electron-positron collisions was analyzed in [3, 9]. No attention has so far 
been given to the production of triply heavy baryons.
  
Investigation of the mechanisms responsible for
the production of multiply heavy hadrons is of interest
from the theoretical point of view since this provides
the possibility of further verifying QCD (more precisely, of obtaining deeper 
insight into this theory).
This involves testing both its perturbative aspect,
which is used to describe the simultaneous production of a few quark pairs, and
QCD-inspired non-perturbative models of bound states. We recall that,
even in apparently obvious cases, the results of QCD
calculations often unexpectedly prove to be at odds
with experimental data, as was, for example, in the
hadronic production of $J / \psi$ particles. On the other
hand, a derivation of theoretical cross-section estimates is of importance from
the point of view of
applications to searches for such particles and investigation into their 
properties. The presence of two or  more heavy quarks in a hadron substantially
affects the properties of its weak decays. At the same time,
the reliability of theoretical predictions is higher for
such hadrons, and this makes it possible to test model
concepts more thoroughly.
   
In the present study, we reveal some features of
$\Omega_{scb}$ baryon production in electron-positron annihilation. The choice 
of a process where the initial
state is purely leptonic was motivated, on one hand,
by the fact that the relevant calculations are simpler
here than in the case of hadronic production and, on
the other hand, by the fact that the case of leptonic
production offers a number of advantages in what is
concerned with a possible experimental observation,
which include favorable background conditions and a
precisely known initial energy.
  
At the quark level, the subprocess $e^-e^+  \rightarrow  s\bar{s}  c\bar{c}  
b\bar{b}$ which is of order $\alpha^2\alpha_s^4$  in conventional perturbation
theory, is associated with the process being considered. In evaluating the 
square of the relevant matrix
element, use is made of the method that was proposed
in [10] and which is referred to as the method of
orthogonal amplitudes (previously, this method was
employed, for example, in [7, 8]). The fusion of the
product $s$, $c$, and $b$ quarks into a $\Omega_{scb}$ baryon is described 
within the standard nonrelativistic approximation [11-13]. A detailed account 
of the technical facet of our calculations is given in Section 2. The
numerical results obtained on the basis of these calculations are discussed in 
Section 3.

\vspace{1.02cm}

{\bf \center \Large 2. Computational procedure}
                   
Our calculations are based on considering the partonic process
\begin{equation} \label{ee}
e^-(p_{e^-}) + e^+(p_{e^+}) \rightarrow 
s(p_{1}, \xi) + c(p_{2}, \zeta) + b(p_{3}, \chi)
+ \bar{s}(p_{4}, \xi ') + \bar{c}(p_{5}, \zeta ') + \bar{b}(p_{6}, \chi '),  
\end{equation}
where the parentheses referring to the colliding particles enclose their 
4-momenta, while the parentheses
referring to the product quarks and antiquarks enclose their 4-momenta and 
color indices. As usual,
we disregard Feynman diagrams featuring the electroweak interaction of the 
quarks involved. In other
words, we consider only those diagrams where the
interaction between the quarks is mediated by gluons.
For this class of diagrams, Fig. 1 shows nine basic diagrams in which quark and
gluon lines are connected
in different ways. Here, six (three) nonequivalent dispositions of the 
$s\bar{s}$, $c\bar{c}$, and  $b\bar{b}$  lines correspond to each
of the first seven (last two) diagrams in Fig. 1 - that
is, diagrams 1-7 (diagrams 8 and 9). Considering
that the annihilation channel may involve either a
photon or a $Z$ boson, we conclude that the total
number of diagrams in question is 96.

The matrix element for the process in (1) can be represented in the form
\begin{equation}
{\cal M}=\frac{g_{s}^{4}g^{2}}{24\cos^{2}_{\theta_{W}}  
(s-M^{2}_{Z}+iM_{Z}\Gamma_{Z})} \varepsilon^{\xi\zeta\chi} 
\varepsilon^{\xi' \zeta' \chi'} A^{Z}_{\xi \xi' \zeta \zeta' \chi \chi'}
-\frac{g_{s}^{4}e^{2} }{6s} \varepsilon^{\xi\zeta\chi}
\varepsilon^{\xi' \zeta' \chi'} 
A^{\gamma}_{\xi \xi' \zeta \zeta' \chi \chi'},
\end{equation}
where
\begin{eqnarray}
A^{Z}_{\xi \xi' \zeta \zeta' \chi \chi'}&=&
B^{s\bar{s} c\bar{c} b\bar{b}}_{\xi \xi' \zeta \zeta' \chi\chi'}+
B^{s\bar{s} b\bar{b} c\bar{c}}_{\xi \xi' \chi \chi' \zeta \zeta'}+
B^{c\bar{c} s\bar{s} b\bar{b}}_{\zeta \zeta' \xi \xi' \chi \chi'}+
B^{c\bar{c} b\bar{b} s\bar{s}}_{\zeta \zeta' \chi \chi' \xi \xi'}+
B^{b\bar{b} s\bar{s} c\bar{c}}_{\chi \chi' \xi \xi' \zeta \zeta'}+
B^{b\bar{b} c\bar{c} s\bar{s}}_{\chi \chi' \zeta \zeta' \xi \xi'},
\\[3mm]
B^{s\bar{s} c\bar{c} b\bar{b}}_{\xi \xi' \zeta \zeta' \chi \chi'} &=& 
\left\{                                                  
[({p}_{2}+{p}_{1}+{p}_{4})^{2}-m_{c}^{2}]^{-1}               
[({p}_{e^{+}}+{p}_{e^{-}}-{p}_{6})^{2}-m_{b}^{2}]^{-1}            
({p}_{1}+{p}_{4})^{-2} \times                                     
          \right.   \\[3mm]
&\times& ({p}_{1}+{p}_{2}+{p}_{4}+{p}_{5})^{-2}              
\bar{u} ({\bf p}_{2}) T^{a}_{\zeta l}  T^{b}_{l \zeta'} \gamma^{\nu}    
(\hat{p}_{2}+\hat{p}_{1}+\hat{p}_{4}+m_{c})                    
\gamma_{\delta}                                                
v (-{\bf p}_{5}) \times   
            \nonumber  \\  
&\times&
\bar{u}  ({\bf p}_{3}) T^{b}_{\chi \chi '}  \gamma^{\delta}                 
(\hat{p}_{e^{+}}+\hat{p}_{e^{-}}-\hat{p}_{6}+m_{b})            
\gamma_{\varepsilon}                                          
(g^{b}_{V}- g^{b}_{A}\gamma_{5})                             
v (-{\bf p}_{6}) +        
            \nonumber  \\
&+&     
[({p}_{5}+{p}_{1}+{p}_{4})^{2}-m_{c}^{2}]^{-1}               
[({p}_{e^{+}}+{p}_{e^{-}}-{p}_{3})^{2}-m_{b}^{2}]^{-1}            
({p}_{1}+{p}_{4})^{-2}   \times                                    
           \nonumber  \\
&\times& ({p}_{1}+{p}_{2}+{p}_{4}+{p}_{5})^{-2}              
\bar{u}   ({\bf p}_{2})           T^{b}_{\zeta l}            
T^{a}_{l \zeta'}                  \gamma_{\delta}             
(-\hat{p}_{5}-\hat{p}_{1}-\hat{p}_{4}+m_{c})                   
\gamma^{\nu}                                                   
v (-{\bf p}_{5})  \times  
            \nonumber  \\  
&\times&
\bar{u}   ({\bf p}_{3})           T^{b}_{\chi \chi '}      
\gamma_{\varepsilon}                 
(-\hat{p}_{e^{+}}-\hat{p}_{e^{-}}+\hat{p}_{3}+m_{b})           
\gamma^{\delta}                                         
(g^{b}_{V}- g^{b}_{A}\gamma_{5})                            
v (-{\bf p}_{6}) +                      
            \nonumber  \\
&+&    
[({p}_{5}+{p}_{1}+{p}_{4})^{2}-m_{c}^{2}]^{-1}              
[({p}_{e^{+}}+{p}_{e^{-}}-{p}_{6})^{2}-m_{b}^{2}]^{-1}           
({p}_{1}+{p}_{4})^{-2} \times                                    
           \nonumber  \\
&\times& ({p}_{1}+{p}_{2}+{p}_{4}+{p}_{5})^{-2}             
\bar{u}   ({\bf p}_{2})           T^{b}_{\zeta l}           
T^{a}_{l \zeta'}                  \gamma_{\delta}            
(-\hat{p}_{5}-\hat{p}_{1}-\hat{p}_{4}+m_{c})                  
\gamma^{\nu}                                                  
v (-{\bf p}_{5})\times    
            \nonumber  \\  
&\times&
\bar{u}   ({\bf p}_{3})           T^{b}_{\chi \chi '}      
\gamma^{\delta}                 
(\hat{p}_{e^{+}}+\hat{p}_{e^{-}}-\hat{p}_{6}+m_{b})           
\gamma_{\varepsilon}                                         
(g^{b}_{V}- g^{b}_{A}\gamma_{5})                            
v (-{\bf p}_{6}) +                      
            \nonumber  \\
&+&     
[({p}_{2}+{p}_{1}+{p}_{4})^{2}-m_{c}^{2}]^{-1}              
[({p}_{e^{+}}+{p}_{e^{-}}-{p}_{3})^{2}-m_{b}^{2}]^{-1}           
({p}_{1}+{p}_{4})^{-2} \times                                    
           \nonumber  \\
&\times& ({p}_{1}+{p}_{2}+{p}_{4}+{p}_{5})^{-2}             
\bar{u}   ({\bf p}_{2})           T^{a}_{\zeta l}           
T^{b}_{l \zeta'}                  \gamma^{\nu}               
(\hat{p}_{2}+\hat{p}_{1}+\hat{p}_{4}+m_{c})                   
\gamma_{\delta}                                               
v (-{\bf p}_{5}) \times  
            \nonumber  \\  
&\times&
\bar{u}   ({\bf p}_{3})           T^{b}_{\chi \chi '}      
\gamma_{\varepsilon}                                          
(-\hat{p}_{e^{+}}-\hat{p}_{e^{-}}+\hat{p}_{3}+m_{b})          
\gamma^{\delta}                                              
(g^{b}_{V}- g^{b}_{A}\gamma_{5})                            
v (-{\bf p}_{6})+                                            
            \nonumber  \\
&+&  
[({p}_{2}+{p}_{5}+{p}_{3})^{2}-m_{b}^{2}]^{-1}              
[({p}_{e^{+}}+{p}_{e^{-}}-{p}_{6})^{2}-m_{b}^{2}]^{-1}           
({p}_{1}+{p}_{4})^{-2} \times                                    
           \nonumber  \\
&\times& ({p}_{2}+{p}_{5})^{-2}                                 
\bar{u}   ({\bf p}_{2})           T^{b}_{\zeta \zeta '}     
\gamma_{\delta} v (-{\bf p}_{5})                            
\bar{u}   ({\bf p}_{3})                                         
T^{b}_{\chi l}                                              
T^{a}_{l \chi'}                  \gamma^{\delta}             
(\hat{p}_{2}+\hat{p}_{5}+\hat{p}_{3}+m_{b})  \times                 
                \nonumber  \\  
&\times&
\gamma^{\nu}                                                  
(\hat{p}_{e^{+}}+\hat{p}_{e^{-}}-\hat{p}_{6}+m_{b})           
\gamma_{\varepsilon}                                         
(g^{b}_{V}- g^{b}_{A}\gamma_{5})                            
v (-{\bf p}_{6}) +                                           
            \nonumber  \\
&+&     
[({p}_{2}+{p}_{5}+{p}_{3})^{2}-m_{b}^{2}]^{-1}              
[({p}_{1}+{p}_{4}+{p}_{6})^{2}-m_{b}^{2}]^{-1}                   
({p}_{1}+{p}_{4})^{-2}\times                                      
           \nonumber  \\
&\times& ({p}_{2}+{p}_{5})^{-2}
\bar{u}   ({\bf p}_{2})           T^{b}_{\zeta \zeta '}     
\gamma_{\delta} v (-{\bf p}_{5})                            
\bar{u}   ({\bf p}_{3})                                         
T^{b}_{\chi l}                                              
T^{a}_{l \chi'}                  \gamma^{\delta}             
(\hat{p}_{2}+\hat{p}_{5}+\hat{p}_{3}+m_{b}) \times                  
                \nonumber  \\  
&\times&
\gamma_{\varepsilon}                                          
(-\hat{p}_{1}-\hat{p}_{4}-\hat{p}_{6}+m_{b})                  
\gamma^{\nu}                                                 
(g^{b}_{V}- g^{b}_{A}\gamma_{5})                            
v (-{\bf p}_{6})+                                            
            \nonumber  \\
&+&      
[({p}_{1}+{p}_{4}+{p}_{6})^{2}-m_{b}^{2}]^{-1}              
[({p}_{e^{+}}+{p}_{e^{-}}-{p}_{3})^{2}-m_{b}^{2}]^{-1}           
({p}_{1}+{p}_{4})^{-2}\times                                      
           \nonumber  \\
&\times& ({p}_{2}+{p}_{5})^{-2}                                
\bar{u}   ({\bf p}_{2})           T^{b}_{\zeta \zeta '}     
\gamma_{\delta} v (-{\bf p}_{5})                            
\bar{u}   ({\bf p}_{3})                                         
T^{b}_{\chi l}                                              
T^{a}_{l \chi'}                  \gamma_{\varepsilon}        
(-\hat{p}_{e^{+}}-\hat{p}_{e^{-}}+\hat{p}_{3}+m_{b}) \times         
                \nonumber  \\  
&\times&
\gamma^{\delta}                                               
(-\hat{p}_{1}-\hat{p}_{4}-\hat{p}_{6}+m_{b})                  
\gamma^{\nu}                                                 
(g^{b}_{V}- g^{b}_{A}\gamma_{5})                            
v (-{\bf p}_{6}) -                                           
            \nonumber  \\
&-&  (i/2)                                                          
[({p}_{e^{+}}+{p}_{e^{-}}-{p}_{6})^{2}-m_{b}^{2}]^{-1}      
({p}_{1}+{p}_{4})^{-2}                                      
({p}_{2}+{p}_{5})^{-2} \times                                     
           \nonumber  \\
&\times& ({p}_{1}+{p}_{2}+{p}_{4}+{p}_{5})^{-2}                    
f^{abd}[(-p_{1}^{\delta}-p_{4}^{\delta}+p_{2}^{\delta}+p_{5}^{\delta})  
g^{\mu\nu} +                                                 
            \nonumber  \\
&+& (-2p_{2}^{\nu}- 2p_{5}^{\nu})  g^{\mu\delta} +
(2p_{1}^{\mu} + 2p_{4}^{\mu})  g^{\nu\delta} ]              
\bar{u}   ({\bf p}_{2})           T^{b}_{\zeta \zeta '}     
\gamma_{\mu} v (-{\bf p}_{5})  \times                             
            \nonumber  \\
&\times& \bar{u}   ({\bf p}_{3})                                
T^{d}_{\chi \chi'}                                          
                                   \gamma_{\delta}             
(\hat{p}_{e^{+}}+\hat{p}_{e^{-}}-\hat{p}_{6}+m_{b})           
\gamma_{\varepsilon}                                         
(g^{b}_{V}- g^{b}_{A}\gamma_{5})                            
v (-{\bf p}_{6}) -                                            
               \nonumber  \\  
&-&   (i/2)
[({p}_{e^{+}}+{p}_{e^{-}}-{p}_{3})^{2}-m_{b}^{2}]^{-1}      
({p}_{1}+{p}_{4})^{-2}                                      
({p}_{2}+{p}_{5})^{-2}  \times                                    
           \nonumber  \\
&\times& ({p}_{1}+{p}_{2}+{p}_{4}+{p}_{5})^{-2}             
f^{abd}[(-p_{1}^{\delta}-p_{4}^{\delta}+p_{2}^{\delta}+p_{5}^{\delta})  
g^{\mu\nu} +                                                 
            \nonumber  \\
&+& (-2p_{2}^{\nu}- 2p_{5}^{\nu})  g^{\mu\delta} +
(2p_{1}^{\mu} + 2p_{4}^{\mu})  g^{\nu\delta} ]              
\bar{u}   ({\bf p}_{2})     T^{b}_{\zeta \zeta '}    
\gamma_{\mu} v (-{\bf p}_{5})\times                               
             \nonumber  \\
&\times& \left. \bar{u}   ({\bf p}_{3})                         
T^{d}_{\chi \chi'}  \gamma_{\varepsilon}        
(-\hat{p}_{e^{+}}-\hat{p}_{e^{-}}+\hat{p}_{3}+m_{b})          
\gamma_{\delta}                                              
(g^{b}_{V}- g^{b}_{A}\gamma_{5})                            
v (-{\bf p}_{6}) \right\} \times                  
            \nonumber  \\
&\times&
\bar{u}   ({\bf p}_{1})           T^{a}_{\xi\xi'}          
\gamma_{\nu}                      v (-{\bf p}_{4})          
\bar{v}(-{\bf p}_{e^{+}})                                    
\gamma^{\varepsilon}                                            
(g^{e}_{V}-g^{e}_{A}\gamma_{5})                              
u ({\bf p}_{e^{-}}),       \nonumber     \label{M}       
\end{eqnarray}
and where the expression for  $A^{\gamma}_{\xi\xi ' \zeta\zeta ' \chi\chi '}$
is obtained from the expression  $A^{Z}_{\xi\xi ' \zeta\zeta ' \chi\chi '}$  by
setting in it $g^{e}_{V}=1$, $g^{e}_{A}=0$, $g^{q}_{V}=Q_{q}$, and
$g^{q}_{A}=0$, with $Q_{q}$ being the electric charge of the quark $q$ 
($q=s$, $c$, $b$) in units of the electron charge $e$. 
Since the number of nonequivalent Feynman diagrams belonging to type 8
or 9 (Fig. 1) and differing from one another only by
a permutation of the $s\bar{s}$, $c\bar{c}$,
and $b\bar{b}$ lines is equal to three and since the quantity  
$A^{Z}_{\xi\xi'\zeta\zeta'\chi\chi'}$ from Eq. (3)
involves six terms that are obtained from one another
by permuting the quantum numbers of $s\bar{s}$, $c\bar{c}$, and $b\bar{b}$ 
quark-antiquark pairs, we introduce, in the eighth
and the ninth term in expression (4), an additional (in
relation to the Feynman formulation) factor of 1/2,
whereupon we arrive at the correct results for this
quantity and for the matrix element in (2).

Let us consider in more detail the color structure
of the matrix element given by Eqs. (2)-(4). Since
any baryon is a color-singlet object, the $scb$ state
that is produced in process (1) must be an $SU(3)_{c}$
singlet that is contained in the tensor product of three $SU(3)_c$
 triplets. It follows that the $scb$ state
must be fully antisymmetric in the color indices of the
quarks; to take this into account, it is necessary to
introduce, in the amplitude of the process, the antisymmetric tensor
$\varepsilon^{\xi'\zeta'\chi'}/\sqrt{6}$, which is normalized to
unity. Since the initial electron-positron state is also
a color singlet, the state of three unbound $\bar{s}$,
$\bar{c}$ and $\bar{b}$ antiquarks accompanying the product baryon must
also be a singlet. We note in passing that the tensor
$\varepsilon^{\xi\zeta\chi}
T^{a}_{\xi\xi'}T^{a}_{\zeta l}T^{b}_{l\zeta'}T^{b}_{\chi\chi'}$
is already fully antisymmetric in its indices  
$\xi', \zeta'$ and $\chi'$ ; therefore, it  not necessary is
$\varepsilon^{\xi'\zeta'\chi'}/\sqrt{6}$
that the projection operator  be explicitly
present there. The presence of this operator is technically useful, however, 
since this makes it possible to
perform summation over color indices at the amplitude level--that is, prior to 
squaring the amplitude.

Further, it is straightforward to prove the identity\footnote{Of three 
indices $a$, $b$ and $d$ corresponding to nonzero values
of the structure constant $f^{abd}$ two are always the numbers of Gell-Mann 
matrices such that they undergo no changes
upon transposition, while the remaining index is associated
with a Gell-Mann matrix that changes sign upon transposition. By simultaneously
replacing the primed indices by unprimed ones and transposing 
the matrices $T^a$, $T^b$ and $T^d$, we arrive at
$$\varepsilon^{\xi\zeta\chi}\varepsilon^{\xi'\zeta'\chi'}
T^{a}_{\xi\xi'} T^{b}_{\zeta\zeta '} T^{d}_{\chi\chi'}=
-\varepsilon^{\xi\zeta\chi} \varepsilon^{\xi'\zeta'\chi'}
T^{a}_{\xi'\xi} T^{b}_{\zeta'\zeta} T^{d}_{\chi'\chi}=
-\varepsilon^{\xi'\zeta'\chi'}\varepsilon^{\xi\zeta\chi}
T^{a}_{\xi\xi'} T^{b}_{\zeta\zeta'} T^{d}_{\chi\chi'},$$
whence we immediately obtain relation (5).}
\begin{equation} \label{zero}
\varepsilon^{\xi\zeta\chi}\varepsilon^{\xi'\zeta'\chi'} 
f^{abd} T^{a}_{\xi\xi'}T^{b}_{\zeta\zeta'}T^{d}_{\chi\chi'}=0,
\end{equation}
from which it follows that the contributions of diagrams involving a 
three-gluon vertex [eighth and
ninth term in Eq. (4)] vanish. At the same time, all
of the remaining terms in Eqs. (2)-(4) have the same
color structure. Summation over color indices yields
\begin{equation}
\frac{1}{6} \varepsilon^{\xi\zeta\chi}\varepsilon^{\xi'\zeta'\chi'} 
T^{a}_{\xi\xi'}T^{a}_{\zeta l}T^{b}_{l\zeta'} T^{b}_{\chi\chi'}=\frac{4}{9}.
\end{equation}
In performing summation over fermion polarizations (with the aid of the REDUCE 
system [14] for analytic calculations), we employed the method
of orthogonal amplitudes. Briefly, the essence of
the method is as follows. Suppose that we have
the quantity $\bar{u}({\bf p}')  R  u({\bf p}'')$, where $u({\bf p}')$ and 
$u({\bf p}'')$ are spinors that obey the Dirac equation, while $R$
is an operator that is expressed in terms of the 
matrices and their contractions with 4-vectors. In
general, this quantity then admits a linear decomposition in terms of four 
orthogonal amplitudes 
$w_{1}=\bar{u}({\bf p}') u({\bf p}'')$, 
$w_{2}=\bar{u}({\bf p}') \hat{K} u({\bf p}'')$,  
$w_{3}=\bar{u}({\bf p}') \hat{Q} u({\bf p}'')$, and
$w_{4}=\bar{u}({\bf p}') \hat{K}\hat{Q} u({\bf p}'')$.
That two different amplitudes are orthogonal implies the vanishing of the
quantity obtained by summing, over the polarizations
of the two spinors, the product of one of these amplitudes and the complex 
conjugate of the other. This is
so if the 4-vectors $K^{\mu}$ and $Q^{\mu}$ are orthogonal to the
4-momenta and $p'^{\mu}$ and $p''^{ \mu}$ to each other - that is
$K_{\mu}p'^{\mu}=0$, $K_{\mu}p''^{\mu}=0$, $Q_{\mu}p'^{ \mu}=0$,
$Q_{\mu}p''^{ \mu}=0$, and $K_{\mu}Q^{\mu}=0$; 
otherwise, the 4-vectors $K^{\mu}$ and  $Q^{\mu}$ are arbitrary.

Let us apply the method of orthogonal amplitudes
to the specific problem at hand. For this purpose, we
introduce the quantities
\begin{equation}
w_{s1}=\bar{u} ({\bf p}_{1}) v(-{\bf p}_{4}), \hspace{0.3cm}  
w_{s2}=\bar{u} ({\bf p}_{1}) \hat{K}_{s} v(-{\bf p}_{4}),
\end{equation}
$$w_{s3}=\bar{u} ({\bf p}_{1})\hat{Q}_{s} v(-{\bf p}_{4}), \hspace{0.3cm}  
w_{s4}=\bar{u} ({\bf p}_{1})\hat{K}_{s} \hat{Q}_{s} v(-{\bf p}_{4}),$$
$$w_{c1}=\bar{u} ({\bf p}_{2}) v(-{\bf p}_{5}),  \hspace{0.3cm}  
w_{c2}=\bar{u} ({\bf p}_{2})\hat{K}_{c} v(-{\bf p}_{5}),$$
$$w_{c3}=\bar{u} ({\bf p}_{2})\hat{Q}_{c} v(-{\bf p}_{5}), \hspace{0.3cm} 
w_{c4}=\bar{u} ({\bf p}_{2})\hat{K}_{c}\hat{Q}_{c} v(-{\bf p}_{5}),$$  
$$w_{b1}=\bar{u} ({\bf p}_{3}) v(-{\bf p}_{6}),  \hspace{0.3cm} 
w_{b2}=\bar{u} ({\bf p}_{3})\hat{K}_{b} v(-{\bf p}_{6}),$$
$$w_{b3}=\bar{u} ({\bf p}_{3})\hat{Q}_{b}  v(-{\bf p}_{6}), \hspace{0.3cm} 
w_{b4}=\bar{u} ({\bf p}_{3})\hat{K}_{b}\hat{Q}_{b} v(-{\bf p}_{6}),$$  
$$w_{e_{1}}=\bar{v} (-{\bf p}_{e^{+}}) \hat{K}_{e} u({\bf p}_{{e}^{-}}), 
\hspace{0.3cm} 
w_{e_{2}}=\bar{v} (-{\bf p}_{e^{+}}) \hat{Q}_{e}  u({\bf p}_{{e}^{-}}),$$
where
\begin{eqnarray}
&
K_{s}^{\mu}=\varepsilon^{\mu\nu\rho\sigma}p_{1\nu}p_{4\rho}a_{s\sigma},   
\hspace{0.3cm}
Q_{s}^{\mu}=\varepsilon^{\mu\nu\rho\sigma}p_{1\nu}p_{4\rho}K_{s\sigma}, &  \\ 
&
K_{c}^{\mu}=\varepsilon^{\mu\nu\rho\sigma}p_{2\nu}p_{5\rho}a_{c\sigma},  
\hspace{0.3cm}  
Q_{c}^{\mu}=\varepsilon^{\mu\nu\rho\sigma}p_{2\nu}p_{5\rho}K_{c\sigma}, &   
\nonumber   \\ 
&
K_{b}^{\mu}=\varepsilon^{\mu\nu\rho\sigma}p_{3\nu}p_{6\rho}a_{b\sigma},   
\hspace{0.3cm} 
Q_{b}^{\mu}=\varepsilon^{\mu\nu\rho\sigma}p_{3\nu}p_{6\rho}K_{b\sigma}, &  
\nonumber   \\ 
&
K_{e}^{\mu}= 
\varepsilon^{\mu\nu\rho\sigma}p_{e^{+}\nu}p_{e^{-}\rho}a_{e\sigma},   
\hspace{0.3cm} 
Q_{e}^{\mu}= 
\varepsilon^{\mu\nu\rho\sigma}p_{e^{+}\nu}p_{e^{-}\rho}K_{e\sigma}, & \nonumber
\end{eqnarray}
the 4-vectors $a_{s\sigma}$, $a_{c\sigma}$, $a_{b\sigma}$, and  
$a_{e\sigma}$  being arbitrary.

Our problem is then described in terms of 128 orthogonal amplitudes of the form
\begin{equation} \label{w}
w_{ijkl}=w_{si}w_{cj}w_{bk}w_{el}, \qquad i,j,k=1,2,3,4, \qquad l=1,2.
\end{equation}

In order to find the coefficients $c_{ijkl}$ in the expansion of the matrix 
element (2) in the amplitudes specified by Eq. (9),
\begin{equation}
{\cal M}=\sum_{i,j,k=1}^{4} \sum_{l=1}^{2} c_{ijkl}  w_{ijkl}, 
\end{equation}
we multiply both sides of this equality by the quantity
$w^{*}_{i'j'k'l'}$, sum the result over the polarizations of all
fermions, and make use of the orthogonality of the
different amplitudes $w_{ijkl}$. Denoting by $|w_{ijkl}|^2$ the
quantity obtained by summing, over the polarizations
of all fermions, the squared modulus of the amplitude $w_{ijkl}$, we have
\begin{equation} 
c_{ijkl}= \{ \sum_{polar.} {\cal M} w^{*}_{ijkl} \} /|w_{ijkl}|^2.
\end{equation}
For the squared modulus of the relevant matrix element, summation over the 
polarizations of product
particles and averaging over the polarizations of colliding particles is 
performed by the formula
\begin{equation}
\overline{|{\cal M}|^{2}}=
\frac{1}{4} \sum_{i,j,k=1}^{4} \sum_{l=1}^{2} |c_{ijkl}|^2 |w_{ijkl}|^2.
\end{equation}

We note that we did not include the quantities
$\bar{v} (-{\bf p}_{e^{+}}) u ({\bf p}_{e^{-}})$ and
$\bar{v} (-{\bf p}_{e^{+}}) \hat{K}_{e} \hat{Q}_{e} u({\bf p}_{e^{-}})$
in the list of basic orthogonal amplitudes in (7), since the corresponding 
coefficients in the expansion of the matrix element [Eqs. (2)-(4)] vanish (this
is because the traces that arise in performing summation over
the polarizations of massless electrons and positrons
involve an odd number of Dirac $\gamma$ matrices).

The question of why it is profitable to employ the
method of orthogonal amplitudes is in order here.
Upon directly squaring the matrix element specified
by Eqs. (2)-(4), we would obtain, with allowance
for the equality in (5), 3570 terms, and an individual
operation of summation over particle polarizations
would correspond to each of these terms. Within the
method of orthogonal amplitudes, we compose one
REDUCE code for evaluating traces and tensor contractions that corresponds to 
84 terms in the quantity ${\cal M}w^{*}_{1111}$,
whereupon we apply text editors (for
example, joe or gedit) to perform obvious substitutions in this code, thereby 
obtaining REDUCE codes
for evaluating all 128 coefficients $c_{ijkl}$. We note that
analytic expressions for 128 coefficients $c_{ijkl}$ occupy 370 Mb.

\vspace{1.02cm}

{\bf \center \Large 3. Numerical results}

In order to describe a bound state of heavy
quarks, we make use of the nonrelativistic approximation [11-13], according to 
which the relative velocities of the quarks in a heavy hadron are assumed to
be low. In the case of $S$-wave states, these velocities
can be set to zero. Accordingly, the velocities of all
three quarks in the final state of process (1) are taken
to be identical, while the momenta of the quarks are
assumed to be proportional to their masses; that is,
\begin{equation} \label{mass}
p_1=(m_s/M)\,p, \quad p_2=(m_c/M)\,p, \quad p_3=(m_b/M)\,p, \quad 
M=m_s+m_c+m_b.
\end{equation}

Concurrently, the six-particle phase space of the final
state of process (1) reduces to the four-particle phase
space of the process
\begin{equation} \label{ee-scb}
e^-(p_{e^-}) + e^+(p_{e^+}) \rightarrow \Omega_{scb}(p)
+ \bar{s}(p_{4}) + \bar{c}(p_{5}) + \bar{b}(p_{6}),
\end{equation}
while the probability of bound-state formation is controlled by the value of 
the baryon wave function at the
origin of coordinates, the only model parameter in this
approach. Eventually, the differential cross section for
process (14) assumes the form
$$d\sigma = \frac{  (2\pi)^{4} \overline{|{\cal  M}|^{2} } }{ 2 s  } \cdot  
\frac{ |\psi(0)|^{2} }{ M^{2} }
\delta^{4}(   p_{e^{-}}+p_{e^{+}}- p_{4}- p_{5}- p_{6}- p  ) \times$$ 
\begin{equation} 
\times \frac{d^{3}p_{4}}{ (2\pi)^{3}  2E_{4}}   \cdot 
\frac{d^{3}p_{5}}{ (2\pi)^{3}  2E_{5}}   \cdot
\frac{d^{3}p_{6}}{ (2\pi)^{3}  2E_{6}}   \cdot
\frac{d^{3}p}{ (2\pi)^{3}  2E} .
\end{equation}

In evaluating the cross sections in question, we
employed codes for integration that enter as ingredients into the CompHEP 
package [15]. As a necessary
test, we first of all made sure that numerical values of
the cross sections are identical for different choices
of the 4-vectors $a_{s\sigma}$, $a_{c\sigma}$, $a_{b\sigma}$ , and 
$a_{e\sigma}$ , which are
involved in the construction of the basic amplitudes.
Having proven this, we prescribed ten iterations for
the cross sections, each involving 100~000 steps of
a Monte Carlo sampling of the integrand. The error
in evaluating the total cross sections was 2.0-2.5 $\%$,
while the error in the differential cross sections was
about 10 $\%$, on average.

Among theoretical uncertainties that affect cross-section values, the choice
of renormalization scale in
the running coupling constant for strong interaction,
the values of the baryon wave function at the origin,
and numerical values of the quark masses are of
greatest importance. In what is concerned with the
quark masses, the results of the calculations are
the most sensitive to the choice of value for the
lightest quark (strange one), because, for some gluon
propagators, the minimal values of the denominators
are $4m_s^2$. To illustrate this dependence, we everywhere
present the results obtained at two values of the
strange-quark mass, $m_s$ = 300 and 500 MeV. The remaining parameters were set 
to the following values:
$m_{c}= 1500$  MeV, $m_{b}=4800$ MeV,
$\alpha=\alpha(M_Z)=1/128.0$, $\alpha_s=\alpha_s(M_Z/2)=0.134$,  and
$\sin^2\theta_{\rm W}=\sin^2\theta_{\rm W}(M_Z)=0.2240$;
the numerical value of the
wave function for the spin-3/2 $\Omega_{scb}$ baryon at zero
relative coordinates of its quarks was borrowed
from [16]:
\begin{equation} 
|\psi(0)|^{2} = 0.90 \cdot 10^{-3} \hspace{0.2cm} \mbox{\rm GeV}^{6} .
\end{equation}  
We note that the change in the characteristic energy scale $\alpha_{s}(\mu )$ 
in the running coupling constant $\alpha_{s}(\mu)$ from $\mu=M_Z/2$ to
$\mu=M_Z$ leads to a change in
the calculated cross sections by a common factor of
$[\alpha_s(M_Z)/\alpha_s(M_Z/2)]^4=0.665$.

For electron-positron collisions at $\sqrt{s}=$      
91.2 GeV the table presents the values of the total
cross sections  $\sigma_{\rm tot}$   and  the forward-backward
asymmetry at the $Z$-boson pole. This asymmetry is defined as
\begin{equation}
A_{FB}=(\sigma_{F}-\sigma_{B})/(\sigma_{F}+\sigma_{B}),
\end{equation}
where $\sigma_{F} (\sigma_{B})$ is the cross section for the production
of $\Omega_{scb}$ baryons traveling in the forward (backward)
direction with respect to the direction of the electron momentum.

{\footnotesize {\bf Table 1.} Features of $\Omega_{scb}$ baryon production in 
electron-positron collisions at the Z-boson pole  }

\vspace{0.35 cm}

\begin{tabular}{ |p{4.7cm}|  p{4.7 cm}|  p{4.7 cm}|} \hline
$m_{s},$  MeV & $\sigma_{\rm tot},$  fb   &  $A_{FB}$         \\ \hline 
300 &  0.0534$\pm$0.0014  & 0.162$\pm$0.024            \\ \hline 
500 &  0.0153$\pm$0.0004  & 0.158$\pm$0.016        \\ \hline 
\end{tabular}

\vspace{0.35 cm}

Figure 2 displays the transverse-momentum ($p_{T}$)
and rapidity ($Y$) distributions of $\Omega_{scb}$ baryons at
the strange-quark-mass values of $m_{s}$ = 300 and
500 MeV. For both values of the mass  $m_{s}$, the
differential cross sections $d\sigma/dp_{T}$ peak at $p_{T}$ values
approximately equal to one-fourth of the total energy
of colliding particles, while the quantities $d\sigma/dY$ peak
at small positive values of the rapidity $Y$.

By using the concept of a fragmentation function,
we can represent our numerical results in a simpler
analytic form that is convenient for phenomenological
applications. It is natural to break down the entire
set of diagrams considered here into three groups
that correspond to the fragmentation of $b$, $c$, and $s$
quarks (in accordance with the flavor of quarks that
are produced at the  $\gamma$/$Z$ vertex). Here, the fragmentation of $b$ 
quarks plays a dominant role, whence it
follows that, to a high precision, we can approximate
the differential cross section as (see, for example, [4]
and the discussion on the treatment of experimental
data on electron-positron annihilation in [17, 18])
\begin{equation}
d\sigma/dz = \sigma_{b\bar{b}} \cdot 
D_{b \rightarrow \Omega_{scb}} (z) , 
\end{equation}
where $\sigma_{b\bar{b}}$ is the total cross section for the process
$e^{-}e^{+} \rightarrow b\bar{b}$, $D_{b \rightarrow \Omega_{scb}} (z)$ 
is the function that describes
the fragmentation of a $b$ quark into an $\Omega_{scb}$ baryon,
and the variable $z$ is expressed in terms of the energy
$E$ of the final hadron and its longitudinal momentum
$p_{\parallel}$ as
\begin{equation}
z = (E+p_{\parallel})/(E+p_{\parallel})_{\rm max}. 
\end{equation}

For reasons of practical convenience, the variable $z$ is
often replaced by the variable $x_{p}=p/p_{\rm max}$   [17-20],
which is close to it, or by  $x_{E}=E/E_{\rm max}$   [21]. The
distinction between these definitions vanishes in the
limit of ultrahigh energies, but it can be sizable under
actual conditions.

Experimental results obtained for electron-positron annihilation are usually 
contrasted against the Peterson fragmentation function [22]
\begin{equation}
D(z) \sim \frac{ 1 }{ z [1-(1/z)-\varepsilon/(1-z)]^{2} }, 
\end{equation}
where $\varepsilon$ is a phenomenological parameter.

If one disregards the aforementioned small asymmetry in the angular 
distribution of $\Omega_{scb}$ baryons
and sets $z \simeq x_{E}$, the relation between the differential
distribution of the cross section with respect to the
transverse momentum of the product baryon and the
fragmentation function assumes the form
\begin{equation}
\frac{d\sigma}{dp_{T}}  = \frac{4 \sigma_{b\bar{b}} p_{T}}{s} 
\int \limits^{1}_{2\sqrt{(p_{T}^{2} + M^{2})/s}}   
\frac{D_{ b \rightarrow \Omega_{scb}} (z) dz}
{\sqrt{[z^{2} - 4 M^{2}/s][z^{2} - 4 (M^{2} + p_{T}^{2} )/s]}}.  
\end{equation}
But if the variable $x_{p}$ is used instead of $z$, it is necessary to set 
$M = 0$ in relation (21).

The conclusions drawn from a comparison of relation (21) with our numerical 
results are as follows:
if $z \simeq x_{E}$ , the parameter values of $\varepsilon$ = 0.098 $\pm$ 0.012
and 0.132 $\pm$ 0.018 correspond to the strange-quark
masses of $m_{s}$ = 300 and 500 MeV, respectively; if
$z \simeq x_{p}$ , the corresponding parameter values are $\varepsilon$ =
0.108  $\pm$  0.016 and 0.147 $\pm$ 0.022.

For the sake of comparison, we present values of
the parameter $\varepsilon$ in the Peterson fragmentation function (20) that 
were obtained in experiments where
electron-positron annihilation was explored at $\sqrt{s}$ =
10 GeV: $\varepsilon=0.23^{+0.09}_{-0.06}$   for $c$-quark fragmentation
into $\Lambda_{c}$ [17]; $\varepsilon$ = 0.29 $\pm$ 0.06 for $c$-quark 
fragmentation into $\Sigma_{c}$ [19];  and $\varepsilon= 0.24 \pm 0.08 $ and 
$\varepsilon=0.23^{+0.09}_{-0.06}$ 
for $c$-quark fragmentation into $\Xi_{c}$ according to the
results obtained in [18] and [20], respectively.

\vspace{0.39cm}                  

{\bf \center Acknowledgments }
 
We are grateful to A.E. Pukhov for assistance in
adapting the problem considered in this article to the CompHEP package.

\newpage

\begin{figure}[ht]
\vspace*{-5.7cm}  \hspace*{-2.9cm}
\includegraphics[scale=0.91]{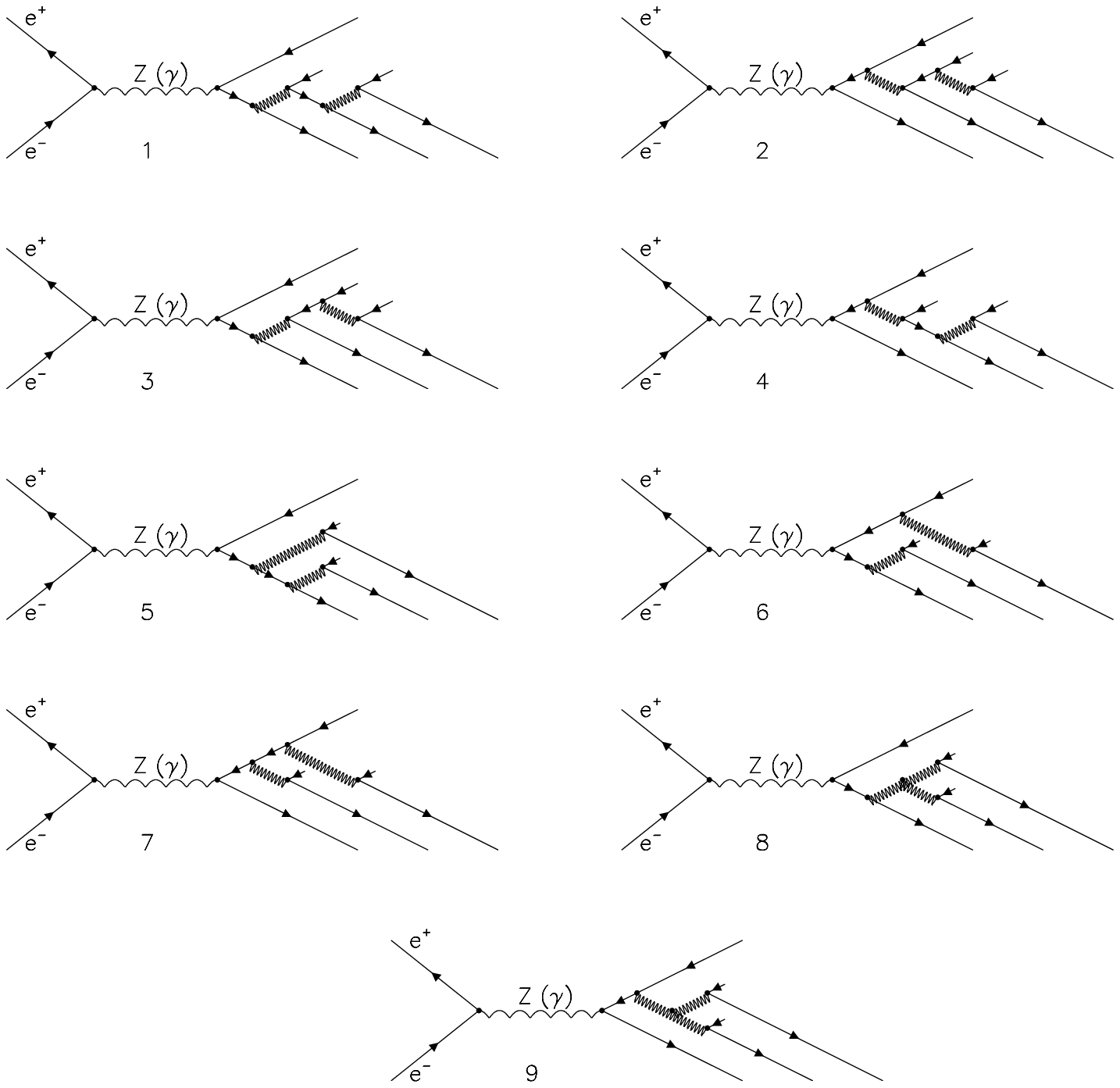}
\vspace*{-7.8cm}\\
{\footnotesize{\bf \hspace*{0.3cm} Fig. 1.} {Basic Feynman diagrams for the 
process $e^{+}+e^{-} \rightarrow s+c+b +\bar{s} +\bar{c}+ \bar{b}$. }}
\end{figure}

\newpage

\begin{figure}[ht]
\centering \includegraphics[width=7.9cm]{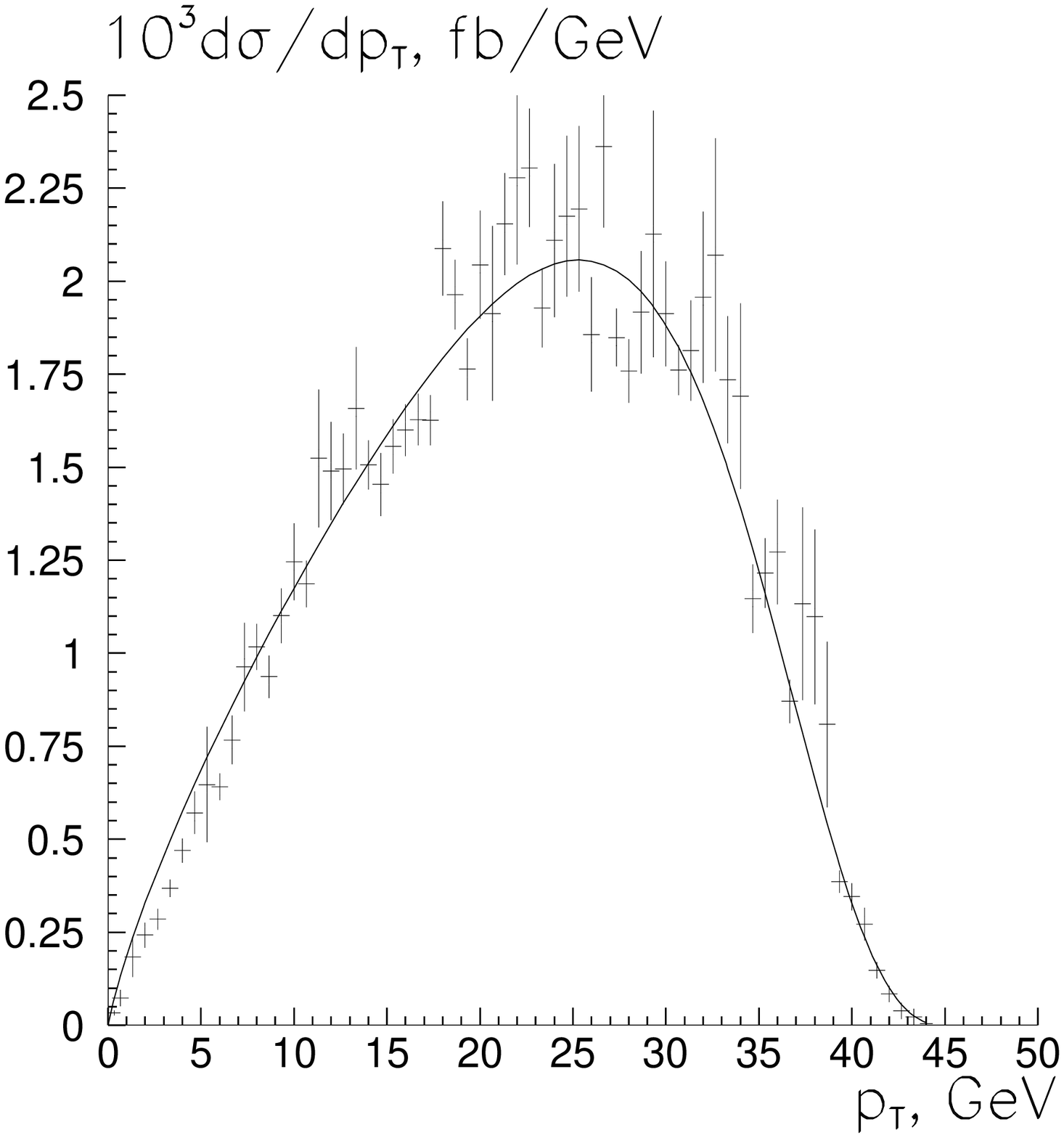} 
\includegraphics[width=7.9cm]{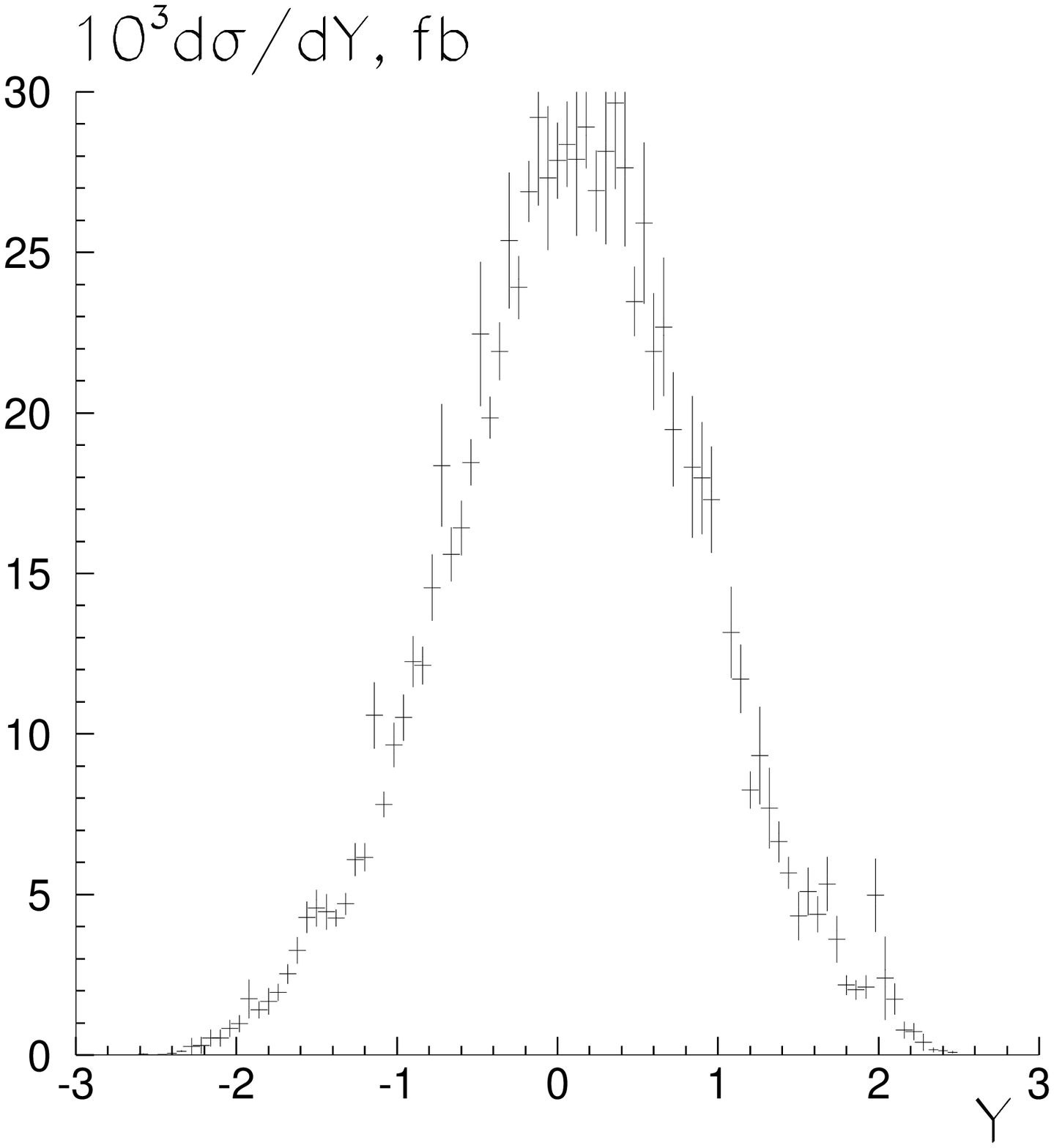}
\end{figure}

\begin{figure}[ht]\vspace*{-0.3cm}
\centering \includegraphics[width=7.9cm]{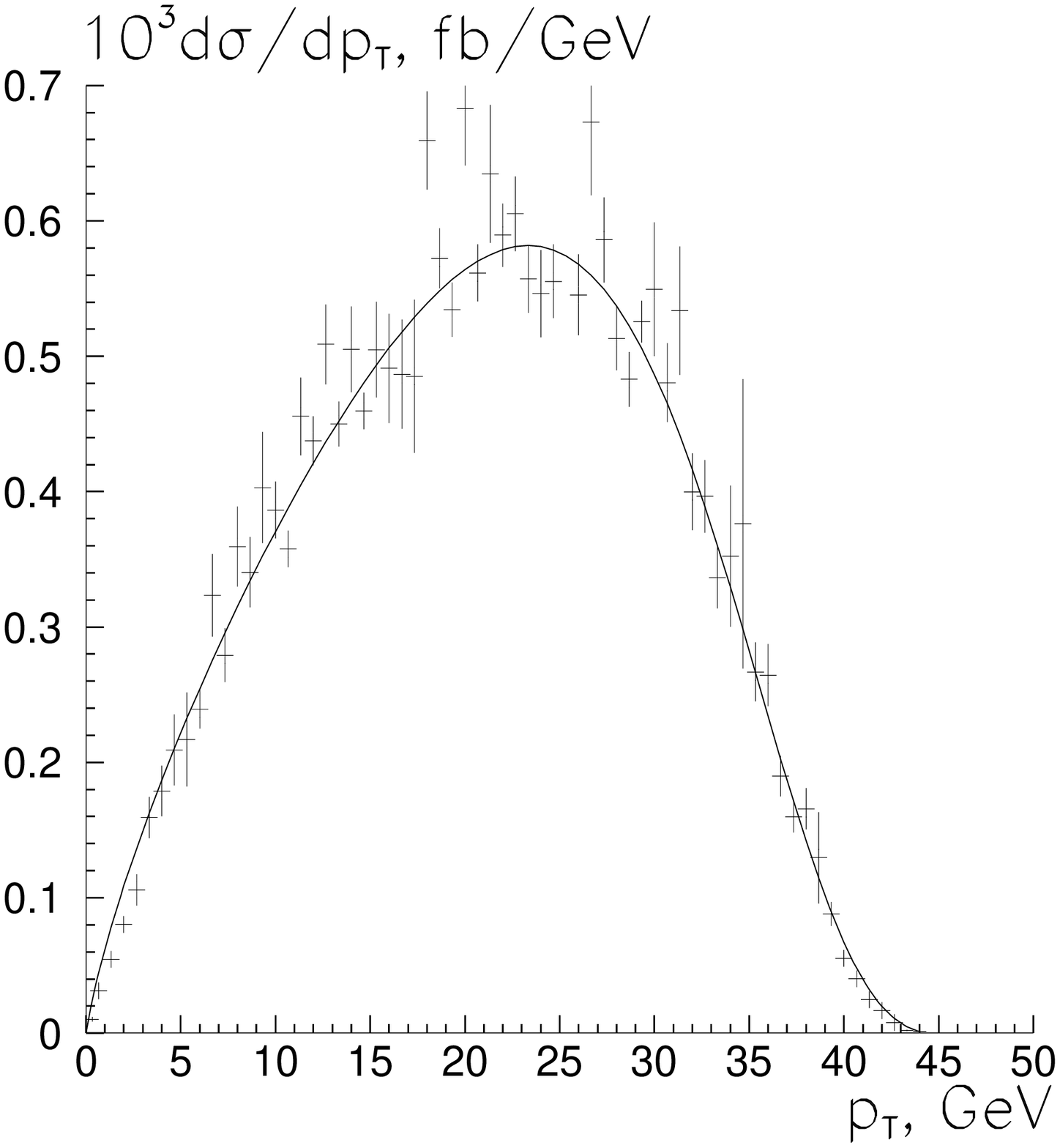} 
\includegraphics[width=7.9cm]{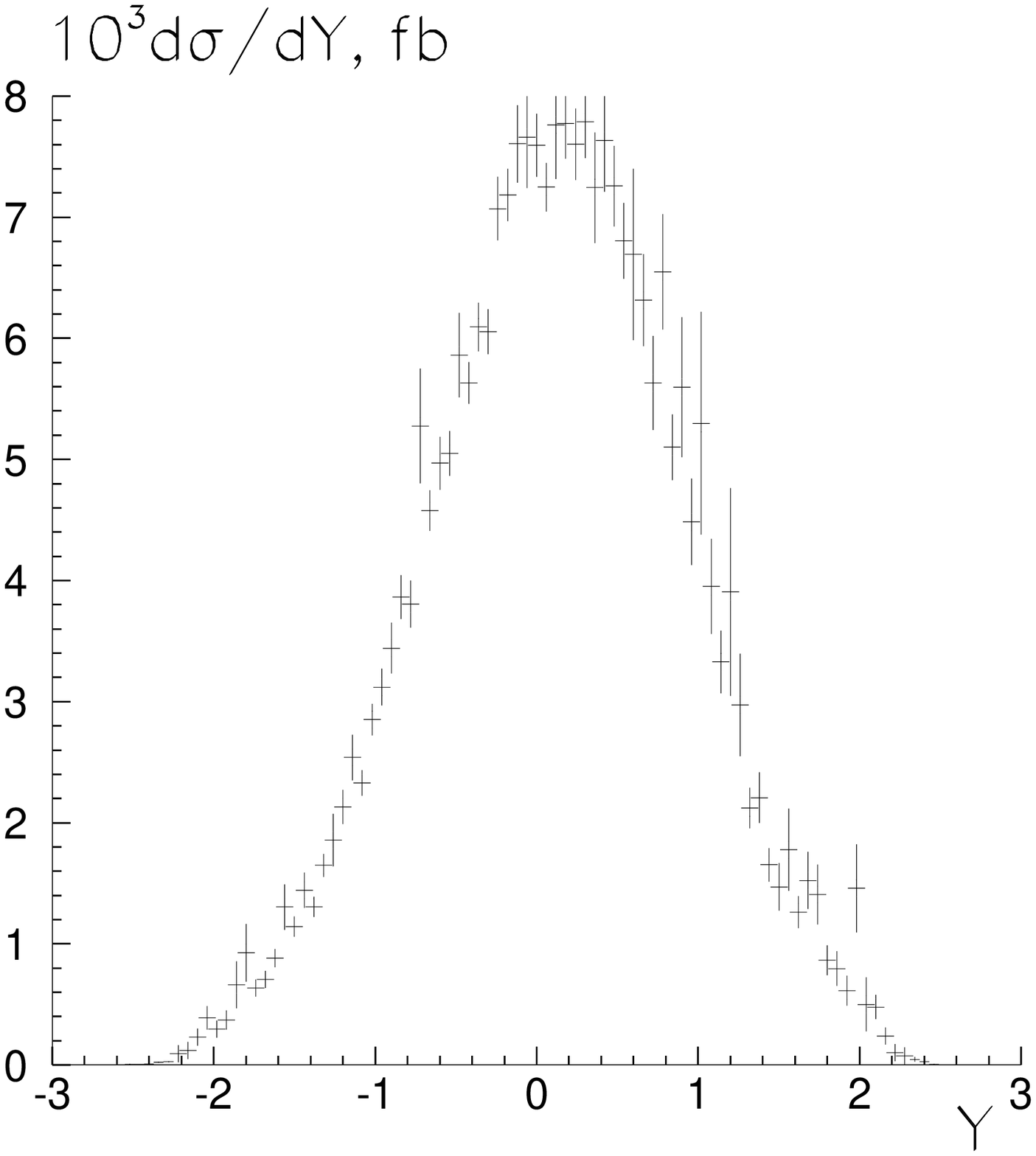}
\end{figure}

\begin{footnotesize}
{\bf Fig. 2.} { Differential distributions of the cross section for
$\Omega_{scb}$ baryon production in electron-positron collisions
at the Z-boson pole with respect to  the transverse momentum $p_{T}$ (top left,
bottom left) and  the rapidity $Y$ (top right, bottom right) at the 
strange-quark-mass values of $m_{s}$ = 300 MeV (top left, top right),  and 
$m_{s}$ = 500 MeV (bottom left, bottom right). Points represent the results of 
our Monte-Carlo calculations. The solid curves correspond to the
calculations by formula (21) with the Peterson fragmentation function whose 
parameter takes the value of
$\varepsilon$ = 0.098 for $m_{s}$ = 300 MeV and the value of 
$\varepsilon$ = 0.132 for $m_{s}$ = 500 MeV. }
\end{footnotesize}


\vspace{1.02cm}

\begin{thebibliography}{99}
%
\bibitem{review1}
V. V. Kiselev and A. K. Likhoded, Usp. Fiz. Nauk  {\bf 172}, 497 (2002).
%
\bibitem{fragm1}
A. Falk, M. Luke, M. Savage, and M. Wise, Phys. Rev. D {\bf 49}, 555 (1994).
%
\bibitem{unbound1}
V.V. Kiselev, A.K. Likhoded, and M.V. Shevlyagin,
              Phys. Lett. B {\bf 332}, 411 (1994).
%
\bibitem{diquark1}
A. V. Berezhnoy, V. V. Kiselev, and A. K. Likhoded, Phys. At. Nucl. {\bf 59}, 
870 (1996).
\bibitem{diquark2}
 A. V. Berezhnoy, V. V. Kiselev, A. K. Likhoded, and
A. I. Onichshenko, Phys. At. Nucl. {\bf 60}, 1875 (1997).
\bibitem{diquark3}
A.V. Bereznoy, V.V. Kiselev, A.K. Likhoded, and A.I. Onishchenko,
             Phys. Rev. D {\bf 57}, 4385 (1998).
\bibitem{diquark4}
S.P. Baranov, Phys. Rev. D {\bf 54}, 3228 (1996).
\bibitem{diquark5}
S.P. Baranov, Phys. Rev. D {\bf 56}, 3046 (1997).  
\bibitem{diquark6}
V.V. Braguta and A.E. Chalov, hep-ph/0005149.
%
\bibitem{Prange}
R.E. Prange, Phys. Rev. {\bf 110}, 240 (1958).
%
\bibitem{CSM1}
C.-H. Chang, Nucl. Phys. B {\bf 172}, 425 (1980).
\bibitem{CSM2}
R. Baier and R. R\"uckl, Phys. Lett. B {\bf 102}, 364 (1981).
\bibitem{CSM3}
E. L. Berger and D. Jones, Phys. Rev. D {\bf 23}, 1521 (1981).
%
\bibitem{REDUCE} A.C. Hearn, Preprint Utah University CP78 Rev. 4/84
(Rand Publ., Utah, 1984).
%
\bibitem{CompHEP}
A. Pukhov {\it et al}., hep-ph/9908288.
%
\bibitem{Psi2}
E. Bagan, H.G. Dosch, P. Godzinsky, S. Narison, and J.-M. Richard,
Z. Phys. C {\bf 64}, 57 (1994).
\bibitem{17}
ARGUS Collab., H. Albrecht {\it et al}., Phys. Lett. B {\bf 207}, 109 (1988).
\bibitem{18}
ARGUS Collab., H. Albrecht {\it et al}., Phys. Lett. B {\bf 247}, 121 (1990).
\bibitem{19}
ARGUS Collab., H. Albrecht {\it et al}., Phys. Lett. B {\bf 211}, 489 (1988).
\bibitem{20}
CLEO Collab., K. W. Edward {\it et al}., Phys. Lett. B {\bf 373}, 261 (1996).
\bibitem{21}
OPAL Collab., G. Alexander {\it et al}., Phys. Lett. B {\bf 364}, 93 (1995).
\bibitem{22}
C. Peterson, D. Schlatter, I. Schmitt, and P. M. Zerwas,  Phys. Rev. D
{\bf 27}, 105 (1983).
%
\end{thebibliography}
\end{document}